\documentclass[
     ,final            
  ]
  {aipproc}

\layoutstyle{8x11single}

\begin{document}

\title{Search for neutrino charged current coherent pion production at SciBooNE}


\author{K.~Hiraide, for the SciBooNE Collaboration}{
  address={Kamioka Observatory, Institute for Cosmic Ray Research, University of Tokyo,
  Gifu 506-1205, Japan}
}

\begin{abstract}
SciBooNE is a neutrino experiment measuring neutrino cross sections on carbon
in the one GeV region. We have performed a search for charged current
coherent pion production from muon neutrinos scattering on carbon,
${\nu_\mu}^{12}{\rm C} \to \mu^{- 12}{\rm C} \pi^+$.
No evidence for coherent pion production is observed.  We set 90\% confidence level
upper limits on the cross section ratio of charged current coherent pion production
to the total charged current cross section at $0.67\times 10^{-2}$ at a mean neutrino
energy of 1.1~GeV and $1.36\times 10^{-2}$ at a mean neutrino energy of 2.2~GeV.
The kinematic distributions of the final data sample are also presented.
\end{abstract}

\maketitle


\section{Introduction}
Neutrino interactions producing single pion form significant backgrounds
for neutrino oscillation searches with a few-GeV neutrino beam,
and thus understanding these processes is essential.
It has been known for years that neutrinos can produce pions by interacting coherently
with the nucleons forming the target nucleus.
Both charged current (CC) and neutral current (NC) coherent modes are
possible, $\nu_\mu A \to \mu^- A \pi^+$ and $\nu_\mu A \to \nu_\mu A \pi^0$, where $A$ is a nucleus.
The outgoing lepton and pion tend to go in the forward direction in the lab frame,
and no nuclear breakup occurs.

There are several experimental measurements in the neutrino energy range
between 1 and 100~GeV, including both the charged current and neutral current modes,
and using both neutrino and antineutrino probes.
These results are well described by the original model of Rein and Sehgal~\cite{Rein:1982pf}.
However, recent results on coherent pion production have induced interest in the neutrino
physics community. The non-existence of charged current coherent pion production
in a 1.3~GeV wide-band neutrino beam has been reported by K2K~\cite{Hasegawa:2005td},
while there exist charged current coherent pion production positive results at higher neutrino energies.
On the other hand, evidence for neutral current coherent pion production in the similar
neutrino energy has been recently reported by MiniBooNE~\cite{AguilarArevalo:2008xs}.

The SciBooNE experiment~\cite{AguilarArevalo:2006se} is designed to measure the neutrino
cross sections on carbon in the one GeV region.
In this paper, we report the first measurement of charged current coherent pion production
on carbon by neutrinos in the SciBooNE experiment, which was published in ref.~\cite{Hiraide:2008eu}.
The kinematic distributions of the final data sample are also presented.

\section{The SciBooNE experiment}
The experiment uses the Booster Neutrino Beam (BNB) at Fermilab.
The primary proton beam, with kinetic energy 8~GeV, is extracted to strike
a 71~cm long, 1~cm diameter beryllium target. Each beam spill consists of 81 bunches of protons,
containing typically $4\times 10^{12}$ protons in a total spill duration of 1.6~$\mu$sec.
The target sits at the upstream end of a magnetic focusing horn that is pulsed with
approximately 170~kA to focus the mesons, primarily $\pi^+$, produced by the $p\rm{-Be}$ interactions.
In a 50 m long decay pipe following the horn, $\pi^+$ decay and produce neutrinos, before
the mesons encounter an absorber. The flux is dominated by muon neutrinos (93\% of total),
with small contributions from muon antineutrinos (6.4\%), and electron neutrinos and
antineutrinos (0.6\% in total). The flux-averaged mean neutrino energy is 0.7 GeV.
When the horn polarity is reversed, $\pi^-$ are focused and hence a predominantly antineutrino
beam is created.

The SciBooNE detector is located 100 m downstream from the neutrino production target.
The detector complex consists of three sub-detectors: a fully active fine grained
scintillator tracking detector (SciBar), an electromagnetic calorimeter (EC) and
a muon range detector (MRD). The SciBar detector consists of 14,336 extruded plastic
scintillator strips, each $1.3\times 2.5\times 300$~cm$^3$. The scintillators are arranged
vertically and horizontally to construct a  $3\times 3\times 1.7$ m$^3$ volume with a total
mass of 15 tons. Each strip is read out by a wavelength-shifting fiber attached
to a 64-channel multi-anode PMT.
Charge and timing information from each MA-PMT is recorded by custom electronics.
The minimum length of a reconstructed track is 8~cm
which corresponds to a proton with momentum of 450~MeV/$c$.
The EC is installed downstream of SciBar, and consists of 32 vertical and 32 horizontal
modules made of scintillating fibers embedded in lead foils. Each module has dimensions
of $4.0\times 8.2\times 262$ cm$^3$, and is read out by two 1'' PMTs on both ends.
The EC has a thickness of $11X_0$ along the beam direction to measure $\pi^0$ emitted
from neutrino interactions and the intrinsic $\nu_e$ contamination.
The energy resolution is $14\%/\sqrt{E\rm{[GeV]}}$.
The MRD is located downstream of the EC in order to measure the momentum of muons up to
1.2 GeV$/c$ with range. It consists of 12 layers of 2''-thick iron plates sandwiched
between layers of 6 mm-thick plastic scintillator planes. The cross sectional area of
each plate is $305\times 274$ cm$^2$. The horizontal and vertical scintillator planes
are arranged alternately, and the total number of scintillators is 362.

The experiment took both neutrino and antineutrino data from June~2007 until August~2008.
In total, $2.64\times 10^{20}$~POT were delivered to the beryllium target during the SciBooNE data
run. After beam and detector quality cuts, $2.52\times 10^{20}$~POT are usable for physics
analyses; $0.99\times 10^{20}$~POT for neutrino data and $1.53\times 10^{20}$~POT for antineutrino
data. Results from the full neutrino data sample are presented in this paper.

\section{Event selection}
The experimental signature of charged current coherent pion production is the existence of
two and only two tracks originating from a common vertex, both consistent with minimum ionizing
particles (a muon and a charged pion).

To identify charged current events, we search for tracks in SciBar matching with a track or hits in the MRD.
Such a track is defined as a SciBar-MRD matched track. The most energetic SciBar-MRD matched
track in any event is considered as the muon candidate.
The matching criteria imposes a muon momentum threshold of 350~MeV/$c$.
The neutrino interaction vertex is reconstructed
as the upstream edge of the muon candidate. We select events whose vertices are in the SciBar fiducial
volume, $2.6 \ {\rm m}\times 2.6 \ {\rm m}\times 1.55 \ {\rm m}$, a total mass of 10.6~tons.
Finally, event timing is required to be within a 2~$\mu$sec beam timing window.
The cosmic-ray background contamination in the beam timing window is only 0.5\%, estimated using
a beam-off timing window.
Approximately 30,000 events are selected as our standard charged current sample,
which is called the SciBar-MRD matched sample.
According to the MC simulation, the selection efficiency and purity of true $\nu_\mu$ charged current events
are 27.9\% and 92.8\%, respectively.
Two subsamples of the SciBar-MRD matched sample are further defined:
the MRD stopped sample and the MRD penetrated sample. Events with the
muon stopping in the MRD are classified as MRD stopped events.
Events with the muon exiting from the downstream end of the MRD are
defined as the MRD penetrated sample, in which we can measure only
a part of the muon momentum. The average neutrino beam energy for
true charged current events in the MRD stopped and MRD penetrated
samples is 1.0 GeV and 2.0 GeV, respectively, enabling a measurement
of charged current coherent pion production at two different mean neutrino energies.

Once the muon candidate and the neutrino interaction vertex are reconstructed,
we search for other tracks originating from the vertex. 
Most events are reconstructed as either one track or two track events.
The two track sample is further divided based on particle identification.
The particle identification variable, Muon Confidence Level (MuCL) is related to the probability
that a particle is a minimum ionizing particle based on the energy deposition.
The probability of misidentification is estimated to be 1.1\% for muons and 12\% for protons.
We first require that the MuCL of the SciBar-MRD matched track is greater
than 0.05 to reject events with a proton penetrating into the MRD.
Then the second track in the event is classified as a pion-like or a
proton-like track with the same MuCL threshold. 

In a charged current resonant pion event, $\nu p\to \mu^- p\pi^+$, the proton is
often not reconstructed due to its low energy, and such an event is
therefore identified as a two track $\mu+\pi$ event. To separate charged current
coherent pion events from charged current resonant pion events, additional protons
with momentum below the tracking threshold are instead detected by
their large energy deposition around the vertex, so-called vertex activity. 
Events with energy deposition greater than 10 MeV are considered to have activity at the
vertex.

Four sub-samples, the one track events, $\mu+p$ events, $\mu+\pi$ events with vertex activity,
and $\mu+\pi$ events without vertex activity are used for constraining
systematic uncertainties in the simulation.
The MC distributions of the square of the four-momentum transfer ($Q^2$) are fitted to the distributions
of the four aforementioned data samples.
The reconstructed $Q^2$ is calculated as
\begin{equation}
  Q^{2}_{\rm rec} = 2 E_{\nu}^{\rm rec} ( E_{\mu} - p_{\mu} \cos \theta_{\mu} ) - m_{\mu}^2
  \label{eq:q2rec}
\end{equation}
where $E_\nu^{\rm rec}$ is the reconstructed neutrino energy calculated by assuming
charged current quasi-elastic (CC-QE) kinematics,
\begin{equation}
  E_{\nu}^{\rm rec} = \frac{1}{2}
  \frac{(m_p^2-m_\mu^2)-(m_n-V)^2+2E_\mu(m_n-V)}{(m_n-V)-E_\mu+p_\mu \cos \theta_\mu}
  \label{eq:enurec}
\end{equation}
where $m_p$ and $m_n$ are the mass of proton and neutron,
respectively, and $V$ is the nuclear potential, which is set to 27~MeV.
The fitting is described in detail in ref.~\cite{Hiraide:2008eu}.

Charged current coherent pion candidates are extracted from the $\mu+\pi$ events which do not have vertex activity.
The sample still contains CC-QE events in which a proton is
misidentified as a minimum ionizing track. We reduce this background by
using kinematic information in the event.  Since the CC-QE
interaction is a two-body interaction, one can predict the proton
direction from the measured muon momentum and muon angle.
For each two-track event, we define an angle called $\Delta\theta_p$
as the angle between the expected proton track and the observed second track directions.
Events with $\Delta\theta_p$ larger than 20 degrees are selected.
With this selection, 48\% of charged current quasi-elastic events in the $\mu+\pi$ sample are
rejected, while 91\% of charged current coherent pion events pass the cut according to the MC simulation.
Further selections are applied in order to separate charged current coherent pion
events from charged current resonant pion events which are the dominant backgrounds
for this analysis.
In the case of charged current coherent pion events, both the muon and pion tracks
are directed forward, and therefore events in which the track angle of the pion
candidate with respect to the beam direction is less than 90~degrees are selected.

\begin{figure}[tbp]
    \begin{tabular}{cc}
      \begin{minipage}{75mm}
        \includegraphics[keepaspectratio=true,height=55mm]{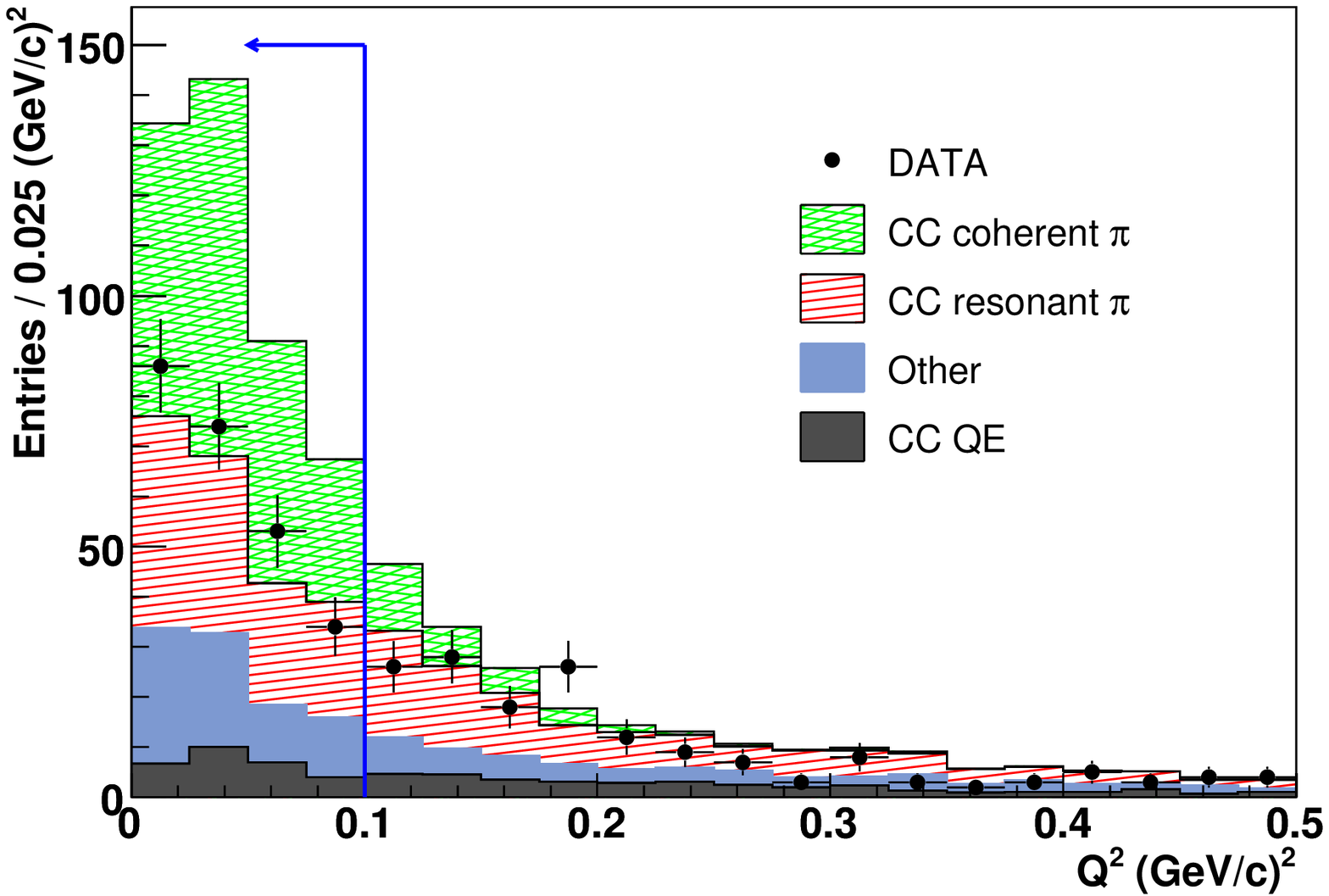}
      \end{minipage}
      &
      \begin{minipage}{75mm}
        \includegraphics[keepaspectratio=true,height=55mm]{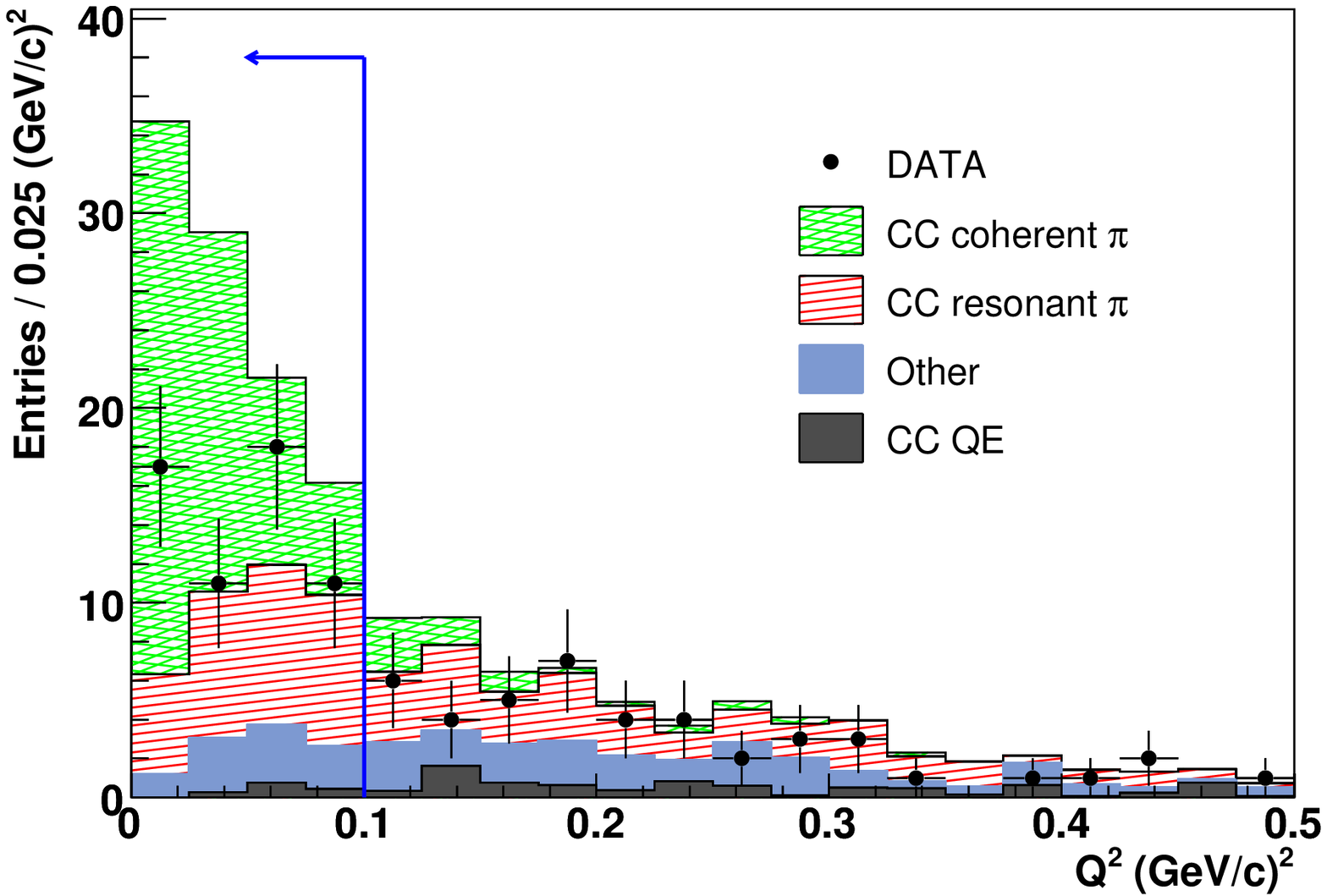}
      \end{minipage}      
    \end{tabular}
  \caption{Reconstructed $Q^2$ for the MRD stopped charged current coherent pion sample (left), and
  the MRD penetrated charged current coherent pion sample (right).}
  \label{fig:q2rec.eps}
\end{figure}

Figure~\ref{fig:q2rec.eps}~(left) shows reconstructed $Q^2$ for
the $\mu+\pi$ events in the MRD stopped sample after the pion track direction cut.
Although a CC-QE interaction is assumed, the $Q^2$ of
charged current coherent pion events is reconstructed with a resolution of
0.016~(GeV/$c$)$^2$ and a shift of -0.024~(GeV/$c$)$^2$ according to the MC simulation.
Finally, events with reconstructed $Q^2$ less than 0.1 (GeV/$c$)$^2$ are selected.
In the signal region, 247 charged current coherent pion candidates are observed,
while the expected number of background events is 228$\pm$12.
The error comes from the errors on the fitting parameters.
The selection efficiency for the signal is estimated to be 10.4\%.
The mean neutrino beam energy for true charged current coherent pion events
in the sample is estimated to be 1.1 GeV after accounting for the effects
of the selection efficiency. 

The same selection is applied to the MRD penetrated sample to extract
charged current coherent pion candidates at higher energy.
Figure~\ref{fig:q2rec.eps}~(right) shows reconstructed $Q^2$ for
the MRD penetrated charged current coherent pion sample.
In the signal region, 57 charged current coherent pion candidates are observed,
while the expected number of background events is 40$\pm$2.2.
The selection efficiency for the signal is estimated to be 3.1\%.
The mean neutrino beam energy for true charged current coherent pion events
in the sample is estimated to be 2.2 GeV. 

\section{$\sigma (\mbox{CC coherent $\pi$})/\sigma (\mbox{CC})$ cross section ratio}
We measure the cross section ratios of charged current coherent pion production to
total charged current interaction with two distinct data samples.
With the MRD stopped sample, the ratio of the charged current coherent
pion production to total charged current cross sections is measured to be
$(0.16\pm 0.17(stat)^{+0.30}_{-0.27}(sys))\times 10^{-2}$.
The result is consistent with the nonexistence of charged current coherent
pion production, and hence we set an upper limit on the cross section ratio
by using the likelihood distribution ($\mathcal{L}$) which is convolved with the
systematic error. We calculate the 90\% confidence level (CL)
upper limit (UL) using the relation
$\int_{0}^{UL} \mathcal{L} dx/\int_{0}^{\infty} \mathcal{L}dx=0.9$
to be
\begin{equation}
	\sigma (\mbox{CC coherent $\pi$})/\sigma (\mbox{CC}) < 0.67\times 10^{-2}
\end{equation}
at a mean neutrino energy of 1.1 GeV.

With the MRD penetrated sample, the cross section ratio is measured to be
$(0.68\pm 0.32(stat)^{+0.39}_{-0.25}(sys))\times 10^{-2}$.
No significant evidence for charged current coherent pion production
is observed, and hence we set an upper limit on the cross section ratio
at 90\% CL:
\begin{equation}
	\sigma (\mbox{CC coherent $\pi$})/\sigma (\mbox{CC}) < 1.36\times 10^{-2}
\end{equation}
at a mean neutrino energy of 2.2 GeV.

According to the Rein-Sehgal model~\cite{Rein:1982pf,Rein:2006di} implemented
in our simulation~\cite{Hayato:2002sd,Mitsuka:2008zz},
the cross section ratio of charged current coherent pion production
to total charged current interactions is expected to be $2.04\times 10^{-2}$.
Our limits correspond to 33\% and 67\% of the prediction at 1.1 GeV and 2.2 GeV, respectively.
Our results are consistent with the K2K result~\cite{Hasegawa:2005td}:
$\sigma (\mbox{CC coherent $\pi$})/\sigma (\mbox{CC})<0.60\times 10^{-2}$ at 90\% CL
measured in a 1.3 GeV wideband neutrino beam.

\begin{figure}[tbp]
    \includegraphics[keepaspectratio=true,height=110mm]{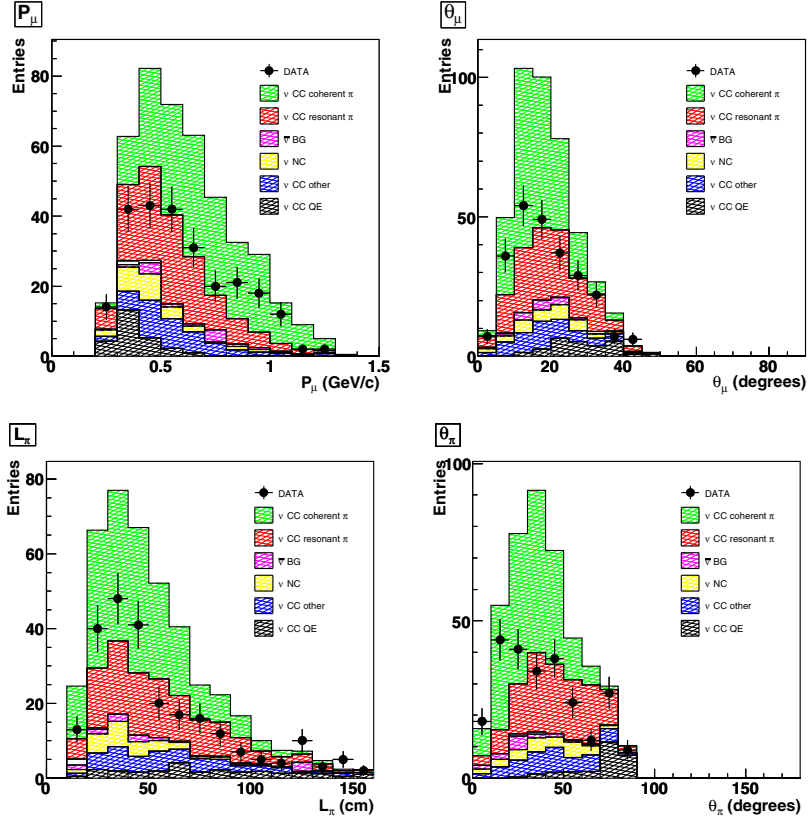}
  \caption{Muon momentum, angle, pion track length, and pion angle distributions
  for the MRD stopped coherent pion sample.}
  \label{fig:kinematics.eps}
\end{figure}

\section{Kinematics distributions}
\begin{figure}[tbp]
    \includegraphics[keepaspectratio=true,height=40mm]{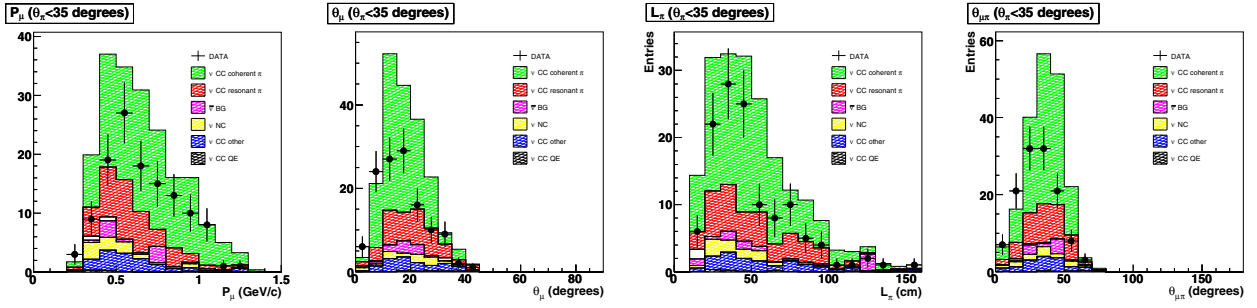}
  \caption{Muon momentum, angle, pion track length, and pion angle distributions
  for the MRD stopped coherent pion events in which pion angle is smaller than 35 degrees.}
  \label{fig:kinematics-le35deg.eps}
\end{figure}

\begin{figure}[tbp]
    \includegraphics[keepaspectratio=true,height=40mm]{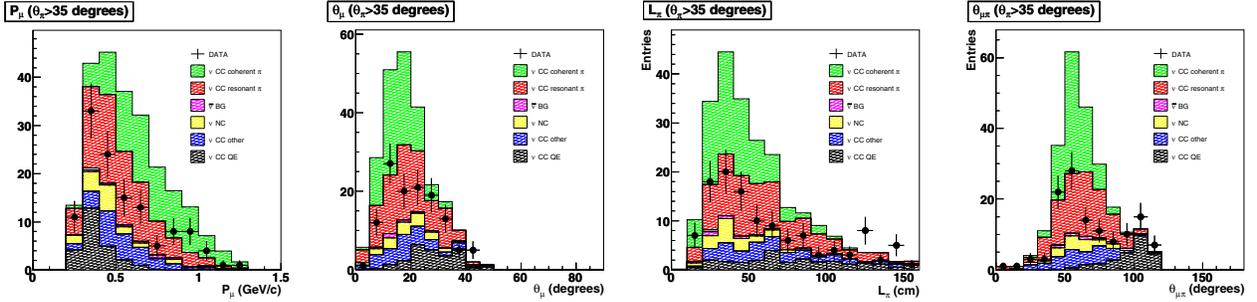}
  \caption{Muon momentum, angle, pion track length, and pion angle distributions
  for the MRD stopped coherent pion events in which pion angle is larger than 35 degrees.}
  \label{fig:kinematics-ge35deg.eps}
\end{figure}

Figure~\ref{fig:kinematics.eps} shows the distributions of muon momentum,
muon angle, pion track length, and pion angle for the MRD stopped coherent pion sample.
The distributions of data in Figure~\ref{fig:kinematics.eps} are basically
in agreement with the background, but the data excess seems to cluster
at a certain kinematic region: at small pion scattered angle, for example.
To investigate the data, the sample is further divided into two sub-samples
based on pion scattering angle.
Figure~\ref{fig:kinematics-le35deg.eps} and Figure~\ref{fig:kinematics-ge35deg.eps}
show the kinematic distributions for the MRD stopped coherent pion events with
$\theta_\pi<35$~degrees and $\theta_\pi>35$~degrees, respectively.
The data in the larger pion angle sample is consistent with background prediction
while there is an enhancement of the data excess in the smaller pion angle sample.
If the data excess is due to charged current coherent pion production,
it suggests that pions from charged current coherent pion production tend to 
go in a more forward direction than the Rein-Sehgal model prediction.
The same test has been performed in SciBooNE's antineutrino charged current coherent pion
sample~\cite{Tanaka:NuInt09}, and we found a similar enhancement of data excess to
that seen in the neutrino data.


\begin{figure}[tbp]
    \includegraphics[keepaspectratio=true,height=50mm]{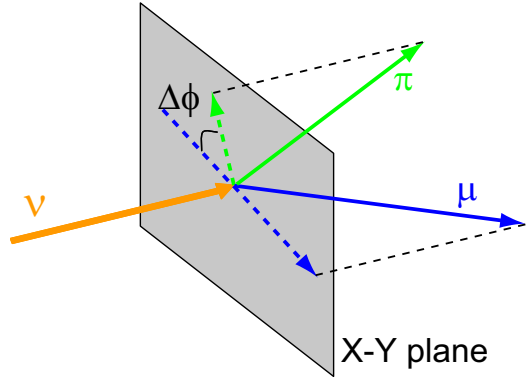}
  \caption{Definition of the kinematic variable $\Delta \phi$.}
  \label{fig:dphi-schematic.eps}
\end{figure}

\begin{figure}[tbp]
    \includegraphics[keepaspectratio=true,height=60mm]{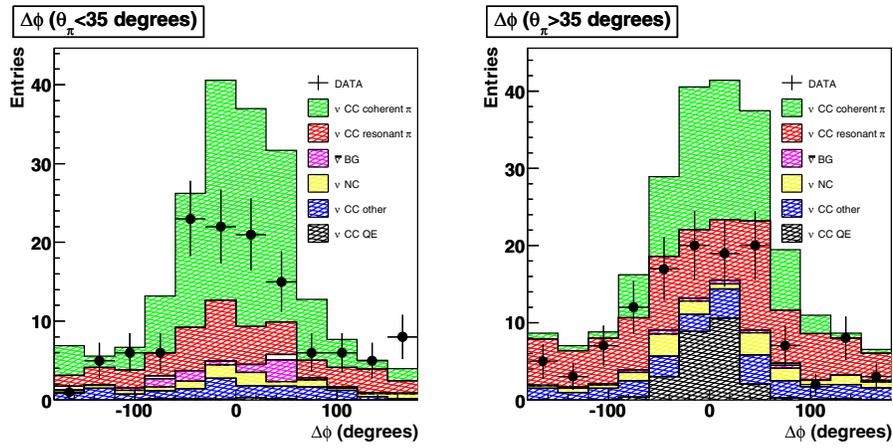}
  \caption{$\Delta \phi$ distributions for two different pion scattered angle
  regions in the MRD stopped coherent pion sample.}
  \label{fig:dphi.eps}
\end{figure}

Another test has been performed to investigate the data.
In the case of charged current coherent pion production,
muon and pion are expected to be emitted back-to-back in the $x$-$y$ plane
because of conservation of momentum. 
Therefore, the kinematic variable called $\Delta \phi$ is defined as shown
in Figure~\ref{fig:dphi-schematic.eps}. The coherent pion events
are expected to distribute around $\Delta \phi=0$.
Figure~\ref{fig:dphi.eps} shows the $\Delta \phi$ distributions for two different
pion scattered angle regions in the MRD stopped coherent pion sample.
In the large pion angle sample, the data and MC distributions agree well.
The charged current quasi-elastic events also distribute around $\Delta \phi=0$
because of two-body interaction.
On the other hand, the data excess is found around $\Delta \phi=0$
in the small pion angle sample. 

These data show that it is important to understand pion kinematics as well as muon kinematics
in order to study charged current pion production.

\section{Summary}
In summary, we have searched for muon neutrino charged current coherent pion production
on carbon in the few-GeV region using the full SciBooNE neutrino data set of
$0.99\times 10^{20}$ POT.
No evidence of charged current coherent pion production is found,
and hence we set 90\% CL upper limits on the cross section ratio
of charged current coherent pion production to total charged current
cross sections at $0.67\times 10^{-2}$ and $1.36\times 10^{-2}$,
at mean neutrino energies of 1.1 GeV and 2.2 GeV, respectively.


\begin{theacknowledgments}
The SciBooNE Collaboration gratefully acknowledge support from various grants,
contracts and fellowships from the MEXT (Japan), the INFN (Italy), the Ministry
of Education and Science and CSIC (Spain), the STFC (UK), and the DOE and NSF (USA).
The author is grateful to the Japan Society for the Promotion of Science for support.
\end{theacknowledgments}

\end{document}